\documentclass[11pt]{article}
\pdfoutput=1
\usepackage{amsmath, amssymb, amsthm}
\newtheorem{theorem}{Theorem}
\newtheorem{corollary}{Corollary}
\newtheorem{lemma}{Lemma}

\theoremstyle{definition}

\usepackage{graphicx}
\usepackage{fullpage}
\usepackage{natbib}

\begin{document}

\title{Representing and decomposing genomic structural variants as balanced integer flows on sequence graphs}
\author{Daniel R. Zerbino$^{1,2*}$, Tracy Ballinger$^{1,2}$, Benedict Paten$^1$, Glenn Hickey$^1$,\\ and David Haussler$^{1,3}$}
\date{}

\maketitle
\bibliographystyle{abbrvnat}

$^1$ Center for Biomolecular Sciences and Engineering, CBSE/ITI, UC Santa Cruz, 1156 High St, Santa Cruz, CA 95064, USA.\\
$^2$ European Molecular Biology Laboratory, European Bioinformatics Institute, Wellcome Trust Genome Campus, Hinxton, Cambridge CB10 1SD, United Kingdom \\
$^3$ Howard Hughes Medical Institute, University of California, Santa Cruz, CA 95064, USA.

$^*$ To whom correspondence should be addressed:\\
Daniel Zerbino\\
European Molecular Biology Laboratory\\ 
European Bioinformatics Institute \\
Wellcome Trust Genome Campus\\
Hinxton\\
Cambridge CB10 1SD \\
United Kingdom\\
Telephone: +44 1223 494 130\\
E-mail: zerbino@ebi.ac.uk

\newpage

\section{Abstract}

The study of genomic variation has provided key insights into the functional role of mutations. Predominantly, studies have focused on single nucleotide variants (SNV), which are relatively easy to detect and can be described with rich mathematical models. However, it has been observed that genomes are highly plastic, and that whole regions can be moved, removed or duplicated in bulk. These structural variants (SV) have been shown to have significant impact on the phenotype, but their study has been held back by the combinatorial complexity of the underlying models. We describe here a general model of structural variation that encompasses both balanced rearrangements and arbitrary copy-numbers variants (CNV). In this model, we show that the space of possible evolutionary histories that explain the structural differences between any two genomes can be sampled ergodically.

\section{Introduction}

Genomic studies, especially in the field of human health, generally do not focus on the majority of bases which are common to all individuals, but instead on the minute differences which are shown to be associated to a variable phenotype \citep{hapmap, 1kg, TCGA, ICGC}. These variants are caused by diverse processes, which modify the genome in different ways. One common task is to find the evolutionary history which most parsimoniously explains the differences between an ancestral genome and a derived genome. 

The best known variants are short single nucleotide variants (SNV) or multiple nucleotide variants (MNVs), which affect at most a handful of consecutive bases. These few bases are substituted, inserted or deleted, without affecting the neighbouring bases or the overall structure of the genome. Especially when only substitutions are taken into consideration, this process can be fully understood using mathematical tools \citep{jukeskantor}: not only is it trivial to describe a parsimonious history that explains the appearance of these variants with a minimum number of mutational events, but posterior likelihoods can be computed across the space of all possible histories.

However, \emph{rearrangement} events sometimes change the overall structure of the genome without changing a base.  Rearrangements can be decomposed into sequences of basic operations, known as double cut and joins (DCJ) \citep{DCJ}. In a DCJ operation the DNA polymer is cleaved at two loci then ligated again, so as to produce a new sequence from the same bases. A DCJ operation can further be decomposed into single cuts or joins (SCJ) \citep{SCJ}. The DCJ operation creates \emph{balanced} rearrangements, i.e. without loss or gain of material. However, coupled with the loss or insertion of detached fragments, these DCJ operations can explain all structural variants, including copy-number variants (CNV) \citep{DCJindels}. These SVs are known to have significant impact on phenotype and health \citep{cnvhealth, cnvcancer}, but the combinatorial complexity of rearrangement models has restricted their study. 

In the absence of CNVs, it is possible to compute a parsimonious history in polynomial time \citep{HP, bergeron, DCJ, SCJ}, but computing its posterior likelihood against all possible histories is computationally prohibitive, as it supposes a random Markov walk across the symmetric group of order $2n$, where $n$ is the number of bases in the genome \citep{durrett}. 

However, in the presence of CNVs, even computing a parsimonious history is difficult, and several teams have approached this problem with slightly different assumptions. Promising results were constructed around a model with a single whole genome duplication and local rearrangements, known as the Genome Halving Problem \citep{elmab1, elmab2, elmab3, elmab4, coloredDB}. Some studies did not allow for duplicated regions \citep{elmab01, Braga}. Others allowed for duplications, considered as independent events on atomic regions, and focusing on the problem of matching segmental copies \citep{DCJindels, nphardsrdd, shaolin, shaomoret, shaolinmoret}. \citet{Bader, baderb} generalized this model further, allowing for larger duplications of contiguous regions along the original genome. 

Other studies, such as \citet{ozeryflato}, extended the SCJ model with an approximate algorithm based on a restricted model of cuts, joins, duplications and deletions. \citet{zeirashamir} demonstrated the NP-hardness of computing an optimal history with fewest duplications. They nonetheless presented a linear time solution to the Genome Halving Problem. 

We described in \citep{avg} the \emph{history graph}. This data structure represents genomes and their evolutionary relationships, allowing for substitutions as well as rearrangements and copy-number changes to occur, much like the trajectory graph defined by \citet{shaolinmoret}. This data structure explicitly represents genomic structure, and assumes that the extant and imputed ancestral genomes are phased, and therefore directly representable as sets of nucleotide sequences. In this model, we were able to compute, between bounds, the number of DCJ events separating two genomes with uneven content, allowing for whole-chromosome duplications and deletions, which are free of cost. This model is extremely general, as it allows segmental duplication and deletion event to occur at any point in the evolutionary history and over any contiguous region at that time point. It also allows duplicate copies to be created then lost during the course of evolution.

However, in practice, it is much easier to estimate the copy-numbers in a sample, using for example shotgun sequencing fragment density with regard to a reference, than it is to construct the underlying genomes. We therefore ask whether it is possible to evaluate the number of rearrangements between a sample and a supposed ancestral genome using copy-numbers and breakpoints alone. By evaluating the number of rearrangements, we are indirectly proposing to simultaneously infer the rearrangement events, and thus assemble the sample's genome, using evolutionary parsimony.

We describe here a general model of structural variation which encompasses both balanced rearrangements and arbitrary copy-numbers variants (CNV). In this model, we show that the difference between any two genomes can be decomposed into a sequence of smaller optimization problems. Mathematically speaking, we represent genomes as points in a $\mathbb{Z}$-module of elements that we call flows, and histories as differences between flows. The overall problem of finding a parsimonious history is then replaced by that of finding an optimal decomposition of the problem, which we address with an ergodic sampling strategy. 

\section{Results}

\subsection{Directed history graphs}

We refer the reader to \citep{avg} for complete definitions of \emph{bidirected history graphs}, \emph{layered histories} and \emph{history epochs} (Figure 1, left side). In the following, we assume that in all layered histories, each segment in a non-root layer is connected to an ancestor in the previous layer.

Given a bidirected history graph $H$ (see \citet{avg}) we construct its \emph{directed history graph} $H'$ as follows: each segment vertex is replaced by a \emph{segment edge} from a \emph{tail} to a \emph{head} vertex, which are distinct. The orientation can be chosen arbitrarily, so long as the DNA label  reflects the sequence that will be read when traversing from tail to head. The bi-directed adjacencies incident on the head side of the segment are connected to the head vertex of the segment, likewise for the tail of the segment. If a branch connects two segments in $H$, a branch similarly connects their head vertices in $H'$, and another branch their tail vertices. 

A directed history graph is trivially equivalent to its bidirected counterpart, so we will use the two concepts interchangeably. Because a vertex in a directed history graph can be incident with at most one adjacency edge and at most one segment edge, each connected component of segment and adjacency edges in $H'$ is a simple path of alternating segment and adjacency edges, which we call a \emph{thread}.

\subsection{Sequence graphs and copy-number}

Henceforth we assume all threads in a layered history are circular, as in Figure 1, thus removing corner cases. The sequence graph $G$ obtained from a layered history is constructed by contracting all the branches of $H$ (Figure 1, right side). Duplicate edges between two vertices of $G$ are merged together. This defines a projection from each edge of $H$ onto the edge of $G$ it is merged into. For a given thread $t$ in $H$, its copy-number weighting on $G$ is the function which assigns to each edge $e$ of $G$ the number of times $t$ traverses a thread edge that projects onto $e$. 

From here onwards, $G$ is assumed to be a sequence graph, and $E$ the set of its edges. Using shotgun sequencing read density and breakpoint analysis, an approximate sequence graph for a thread-structured genome is obtained far more easily than the genome itself, motivating the analysis of such structures.

\begin{figure}[htbp]
\begin{center}
\includegraphics[width=13cm]{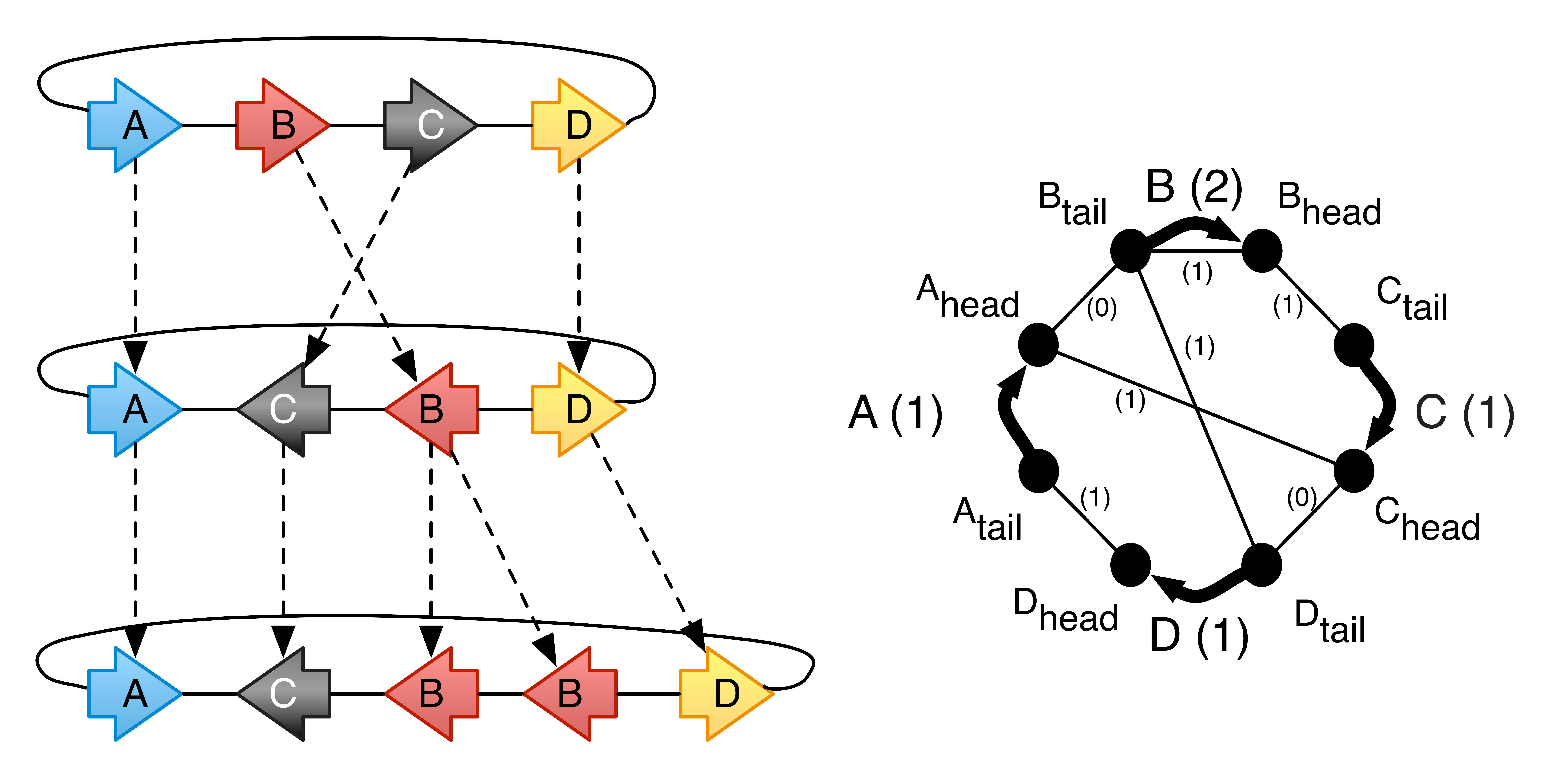}
\caption{Left, a layered directed history graph, where the segments are represented as arrows, thus indicating their head and tail vertices. Right, the corresponding sequence graph, where the segments are represented as thick, curved lines, and the adjacencies as thin straight lines. The copy number weighting, taken from the bottom thread of the layered history graph on the left, is indicated between parentheses for the segment and adjacency edges.}
\label{AncestralSequenceGraph}
\end{center}
\end{figure}

\subsection{Flow functions}

We examine the properties of copy-number weightings on sequence graphs. The \emph{edge space} $\mathbb{R}^E$ denotes the set of real-valued weightings on the set of edges $E$. $\mathbb{R}^E$ is a vector space isomorphic to $ \mathbb{R}^{|E|}$.

A \emph{flow} on $G$ is a weighting in $\mathbb{R}^E$ such that at each vertex the total weight of segment edge incidences equals the total weight of adjacency edge incidences. We call this the \emph{balance condition}. It is essential to distinguish between an \emph{edge} and an \emph{edge incidence}. In particular, if an adjacency edge connects a vertex $v$ to itself, its weight is contributed twice to $v$. Let $\mathcal{F}(G)$ denote the set of all flows on $G$. Because the balance condition is conserved by addition and multiplication by a real scalar, $\mathcal{F}(G)$ is a subspace of $\mathbb{R}^E$. 

Let $\mathcal{F}_{\mathbb{Z}}(G)$ denote the set of integer-valued flows on $G$ and $\mathcal{F}_{\mathbb{Z}}^+(G)$ the cone of non-negative integer flows. $\mathcal{F}_{\mathbb{Z}}(G)$ is a $\mathbb{Z}$-modular lattice of the vector space $\mathcal{F}(G)$. 

\subsection{Copy-number and flow functions}

\begin{lemma}
The copy-number of a circular thread is necessarily a flow in $\mathcal{F}_{\mathbb{Z}}^+(G)$.
\end{lemma}

\begin{proof}
A circular thread describes a cycle that alternates adjacency and segment edges. 
\end{proof}

In a layered history graph $H$, the \emph{flow} of a layer is the sum of copy-number weightings of the threads in that layer. The \emph{flow sequence}  $s(H)$ of a layered history $H$ is the sequence of its layer flows, and conversely $H$ is called a \emph{realization} of $s(H)$. 

\begin{figure}[htbp]
\begin{center}
\includegraphics[width=13cm]{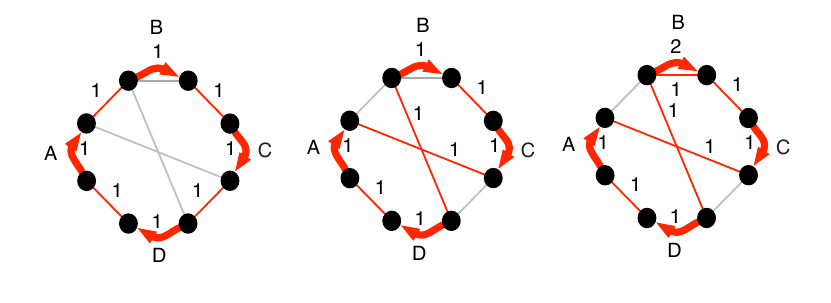}
\caption{The flow sequence associated to the layered history graph in Figure \ref{AncestralSequenceGraph}}
\label{FlowSequence}
\end{center}
\end{figure}

A \emph{valid flow sequence} is a sequence of non-negative integer flows $(f_1,\ldots,f_k)$ in $\mathcal{F}_{\mathbb{Z}}^+(G)^k$ such that for any segment edge $e$, if there exists $i$ such that $f_i(e)=0$ then $f_j(e)=0$ for all $i \le j \le k$. In addition, for every segment edge $e$, $f_1(e)=1$. This ensures that if the number of copies of a segment falls to zero, then it can not be recreated in any subsequent stage of the history, and that at the start of the flow sequence there is exactly one copy of each segment.

A \emph{cycle traversal} is a closed walk through the graph, possibly with edge re-use, with an explicit starting vertex. A \emph{cycle} is an equivalence class of cycle traversals defined by a common circular permutation of edges and its reverse.

\begin{lemma}
Any layered history has a valid flow sequence, and any valid flow sequence has a realization as a layered history.
\end{lemma}

\begin{proof}
The first part follows easily from Lemma 1. To prove the second part, we first decompose each flow in the sequence into a set of threads. To decompose a flow of $\mathcal{F}_{\mathbb{Z}}^+(G)$ into a sum of circular thread flows, we color the edges of $G$ that have non-zero weight such that segments edge are green, and adjacencies orange. Edges with weight 0 are removed. Each edge with weight of absolute value $n > 1$ is replaced by $n$ identical edges of weight 1, creating a multigraph $G'$. This defines a trivial mapping from the edges of $G'$ to those of $G$. By construction, $G'$ is a balanced bi-edge-colored graph (as defined by \citet{PevznerBiColored}), therefore it is possible to find an edge-covering composed of color-alternating cycles. This can be done in polynomial time by a procedure akin to the typical greedy construction on balanced non-colored graphs \citep{Kotzig, PevznerBiColored}. By the mapping defined above, we defined a set $C$ of cycles in $G$ that alternate between segment and adjacency edges. By construction, each edge of $G$ has as many cycle edges mapping to it as its flow weight.

We now demonstrate by induction that we can construct a history graph $H$ from this set of cycle decompositions. If the flow sequence has length 1, we simply construct threads that correspond to the cycles in $C$. Because each segment has a flow of exactly 1, there is a bijection between the newly created segments and the segments of $G$. Now let us assume the property demonstrated for all flow sequences of length $k$ or less. Given a flow sequence $(f_1,\ldots,f_{k+1})$, we first construct a realisation $H'$ of $(f_1,\ldots,f_{k})$. We then choose for every segment edge $e$ in $G$ for which $f_k(e)$ is strictly positive a segment edge in the leaf layer of $H'$ that maps to it. We create under this chosen segment copy $f_{k+1}(e)$ novel descendant segment edges, and assign them an arbitrary order. We  decompose $f_{k+1}$ into a set of cycles in $G$ as described above. For each cycle (picked in any order), we pick a random traversal and create a thread, greedily using up available segments. Once connected by adjacencies, these segments form a thread. By construction, each edge $e$ is visited as many times in the set of cycles as newly created segments map to it, so this process is guaranteed to use up all the newly created segments. By construction, the total flow of these threads is equal to $f_{k+1}$. The history graph thus extended is a realisation of $(f_1,\ldots,f_{k+1})$.
\end{proof}

In \citep{avg}, we defined the \emph{minimum rearrangement} cost of a layered history graph as the minimum number of DCJ operations it would require to account for all the changes that occur in that layered history graph, assuming that whole chromosome duplications and deletions are free of cost. We demonstrated that this cost is NP-hard to compute, but provided upper and lower bounds for it that can be computed in polynomial time. 

In particular, for any directed history graph G, we assign to each vertex $v$ a lifting ancestor $A(v)$ that is its most recent ancestor with an adjacency edge incidence, else it is a new artificial root node if no ancestor of $v$ has an adjacency edge incidence. By adding for each adjacency edge $(u,v)$ a lifted edge $(A(u),A(v))$ we obtain a lifted graph $L(G)$. A lifted edge is trivial if it corresponds to an existing adjacency edge, else it is non-trivial. By removing the segment edges  and branches from $L(G)$, we obtain the module graph $M(G)$, whose connected components are called modules. As demonstrated in \citep{avg}, a lower bound on the rearrangement cost of a directed history graph G is
$$r_l(G)=\sum_{M \in M(G)}{\left( \left \lceil \frac{V_M}{2} \right \rceil -1 \right)}$$
where the sum is over the modules in $M(G)$, and for each module $M$, $V_M$ is the number of its vertices, and an upper bound on the rearrangement cost of $G$ is the number of non-trivial lifted adjacency edges in $L(G)$ minus the number of simple modules, i.e. modules in $M(G)$ in which every vertex has exactly one incident non-trivial lifted adjacency edge. 

The lower bound is closely related to earlier results by \citet{DCJ} and \citet{Bader}.

By extension, we define the minimum rearrangement cost of a valid flow sequence as the minimum rearrangement cost across all of its realizations, and seek to demonstrate the existence of tractable bounds on this value. 

Within a flow sequence $s=(f_1,\ldots,f_k) \in {\mathcal{F}_{\mathbb{Z}}^+(G)}^k, k \ge 2$, for any index $1\le i < k$, the pair $(f_i, f_{i+1})$ is called a \emph{flow transition}. Its \emph{lower complexity} $\mathcal{C}^l_{f_i,f_{i+1}}$ is defined as:
$$\mathcal{C}^l_{f_i,f_{i+1}} = |V| - p - |C|$$
Where:
\begin{itemize} 
\item $V$ is the set of vertices of $G$.
\item $p$ is the number of adjacency edges where $f_i>0$.
\item $C$ is the set of \emph{active components}, \emph{i.e.} connected components of vertices and adjacency edges where $f_i+f_{i+1} > 0$
\end{itemize}

\begin{theorem}
\label{lower}
Given a layered history $H$, decomposed into bilayered history graphs $(g_1,\ldots,g_{k-1})$, with flow sequence $(f_1,\ldots,f_k)$, and any index $i < k$,
$$\mathcal{C}^l_{f_i,f_{i+1}} \le M(g_i)$$ 
\end{theorem}

\begin{proof}
We will prove this by induction on the sequence length $k$. For $k=2$, by definition of valid flow sequences, $f_1$ is equal to 1 on all segment edges, and each vertex is incident on exactly one segment edge, thus every vertex is incident on exactly one adjacency where $f_1 > 0$. This means that in each active component $c \in C$, the number of vertices $v_c$ is exactly twice the number $p_c$ of adjacencies of $c$ where $f_1 > 0$. In the unique bilayered graph we therefore have:
$$|V|-p-|C| = \sum_{c \in C}(v_c-p_c-1) = \sum_{c \in C} \left( \left \lceil \frac{v_c}{2} \right \rceil - 1 \right)$$
The lower complexity is therefore a reformulation of the bound given above.

If $k>2$, to prove the theorem for the flow sequence $(f_1, \ldots, f_k)$ on $(g_1, \ldots, g_{k-1})$, we remove $g_1$ from $H$, and obtain a reduced history graph $H'$ composed of bi-layered sequence graphs $(g_2,\ldots,g_{k-1})$, and its sequence graph $G'$. The sequence graph $G'$ is slightly different from $G$, in that each segment edge of $G'$ has a copy-number of 1 in $f_2$, whereas each segment edge of $G$ has a copy-number of 1 in $f_1$. However, there exists a mapping from the vertices of the second layer of $H$ to the first, implicitly defining a mapping $\Phi$ from the vertices of $G'$ to the vertices of $G$ such that for any edge $e$ in $G$ and any $i \ge 2$, $f_i(e) = \sum_{e' : \Phi(e')=e}f_i'(e')$. On $G'$ we compute the flow sequence of $H'$, $s'=(f'_2, \ldots, f'_{k})$. These flows are all in $\mathcal{F}^+_\mathbb{Z}(G')$

By the inductive hypothesis, the rearrangement cost of $g_{i+1}$ is greater than $\mathcal{C}^l_{f'_i,f'_{i+1}}$, as computed on $G'$. We now compare $\mathcal{C}^l_{f'_i,f'_{i+1}}$ to $\mathcal{C}^l_{f_i,f_{i+1}}$, which is computed on $G$. For simplicity, we create a sequence of  graphs $(G_1=G',\ldots,G_q=G)$, decomposing $\Phi$ into a sequence of projections such that at each step of the sequence, exactly two vertices are merged (note these are not necessarily valid sequence graphs). Edge labels (segment or adjacency) are preserved, and two edges between the same vertices and having the same label are merged together. Projecting $f_i$ and $f_{i+1}$ across this sequence of graphs, we compute $\mathcal{C}^l$ at each step. At each step, the number of vertices $V$ decreases by 1. If the vertices belong to two different active components, the number of active components decreases by 1 and $p$ remains unchanged. Otherwise, the number of active components remains unchanged and $p$ decreases by at most 1, as two edges could be merged into one. The sequence of values of $\mathcal{C}^l$ therefore decreases, hence $\mathcal{C}^l_{f_{i},f_{i+1}} \le \mathcal{C}^l_{f'_{i},f'_{i+1}} \le M(g_{i+1})$. 
\end{proof}

The  \emph{upper complexity} $\mathcal{C}^u_{f_i,f_{i+1}}$ is defined as:
$$\mathcal{C}^u_{f_i,f_{i+1}} = S_i + I_i - l_i$$
Where:
\begin{itemize} 
\item The \emph{duplication count} $D_i$ is a weighting equal to $min(f_i, f_{i+1}-f_i)$ on each edge of $G$. In other words, it is the number of additional copies of each edge in $f_{i+1}$ beyond that in $f_i$, up to a maximum of doubling the number of copies. 
\item The \emph{supra-duplication count} $S_i$ is the sum across all adjacency edges of $max((f_{i+1} - f_i) - D_i, 0)=max(f_{i+1}-2f_i, 0)$.
\item The \emph{de novo edge count} $N_i$ is equal to $f_{i+1}(e)$ if ${f_i}(e) = 0$, else $0$.
\item For any vertex $v$  we denote by  $\mathcal{D}_v$ the sum of $D_i$ on all adjacencies incident on $v$ and by $\mathcal{N}_v$ the sum of $N_i$ on all adjacencies incident on $v$. 
\item The \emph{imbalance} $i_s$ of a segment $s$ between two vertices $a$ and $b$ is equal to $max(\mathcal{D}_a-(\mathcal{D}_b + \mathcal{N}_b), \mathcal{D}_b-(\mathcal{D}_a + \mathcal{N}_a),0)$
\item $I_i$ is the sum of imbalances across all segment edges.
\item $l_i$ is the number of \emph{perfect components}, where a perfect component is a component $c \in C$ such that every vertex of $c$ is incident upon exactly one adjacency $e$ with $f_i(e) = 0$ and $f_{i+1}(e) = 1$, and exactly one  adjacency $e'$ with $f_i(e') > 0$.
\end{itemize}

\begin{figure}[htbp]
\begin{center}
\includegraphics[width=5cm]{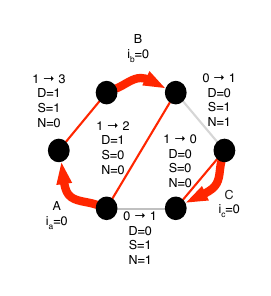}
\caption{Computing the upper bound on a sequence graph. Note the existence of a perfect component connecting segment $C$ to the rest of the sequence graph. In total, we obtain $S=3$, $I=0$, $l=1$, hence an upper bound of 2.}
\label{complexity}
\end{center}
\end{figure}

\begin{theorem}
\label{upper}
Given a flow sequence $s=(f_1,\ldots,f_k) \in {\mathcal{F}_{\mathbb{Z}}^+(G)}^k$, and an index $i < k$, it is possible to find a realisation $H$ with bilayered subgraphs $(g_1, \ldots, g_{k-1})$ such that: 
$$M(g_i) \le \mathcal{C}^u_{f_i,f_{i+1}}$$
\end{theorem}

\begin{proof}
We will construct $g_i$ directly. For every segment edge $s$ of $G$, we create $f_i(s)$ top layer segments, \emph{images} of $s$. In the following, each vertex in $g_i$ is assigned an adjacency to a partner vertex in $G$, on the assumption that our construction algorithm determines independently which specific image in $g_i$ it is connected to.

We start by creating the segment edges of $g_i$. On both ends of a segment $s$, we compute $\mathcal{D} + \mathcal{N}$ and choose the smaller value $m_s$. We select $m_s$ images of $s$ to be marked for \emph{duplication}. The duplicated images of $s$ are assigned 2 descendants each. If $f_{i}(s)=m_s$, an additional $f_{i+1}(s) - 2m_s$ descendant segments are assigned at random to the duplicated segments, otherwise the $f_{i}(s)-m_s$ non-duplicated segments copies are assigned $f_{i+1}(s) - 2m_s$ descendant segments as evenly as possible, such that the difference in number of descendants between any two of those segments is never greater than 1. At the end of this process, segment copies with 0 descendants are marked for \emph{deletion}. Segments which are marked neither for deletion nor duplication are \emph{undetermined}.

Let $d$ be the weighting on $G$ equal to $min(0, f_i-f_{i+1})$, i.e. the number of images of an edge of $G$ which are lost between the two layers of $g_i$. By extension, for any vertex $v$, $d_v$ is the sum of $d$ across all adjacencies incident on $v$.

We then connect the top segments. For every adjacency $e$ of $G$ with $d(e)>0$ we create $d(e)$ images in $g_i$ marked for deletion, attached in priority to segments also marked for deletion. By conservation of flow balance we obtain, $s$ being a segment incident on vertex $v$ on $G$:
$$f_{i+1}(s) - f_i(s) = \mathcal{D}_v + \mathcal{S}_v - d_v$$
The number of segment images marked for deletion is therefore:
$$f_{i}(s) - (f_{i+1}(s) - m_s) \le f_{i}(s) - f_{i+1}(s) + \mathcal{D}_v + \mathcal{N}_v \le f_{i}(s) - f_{i+1}(s) + \mathcal{D}_v + \mathcal{S}_v = d_v$$
Hence all top layer segments marked for deletion can be connected to adjacencies marked for deletion. For every adjacency edge $e$ with a duplication count $D(e)>0$, we create $D(e)$ \emph{duplicated} images of this edge, connected in priority to duplicated copies of $s$, then to undetermined segments. We already connected all segments marked for deletion, hence duplicated adjacencies are connected exclusively to duplicated or undetermined segments. For every adjacency edge $e$ with no flow change in $G$, we create $f_i(e)$ images of this edge connected to the remaining unattached segments. 

We finally connect the bottom segments. If two duplicated segments are connected by a duplicated edge, then two of their descendant segments are connected by adjacencies, thus creating two trivial lifting adjacencies in $g_i$. Otherwise, for each segment which is incident on an adjacency not marked for deletion, it necessarily has at least one descendant segment, which is is connected by an adjacency to a descendant of its partner, thus creating one trivial lifting adjacency in $g_i$. All remaining bottom segments are connected at random, conditional on respecting the adjacency counts specified by $f_{i+1}$.

We now evaluate the upper bound rearrangement complexity of the corresponding DNA history graph as quoted above from \citep{avg}. By construction, deleted adjacencies and adjacencies with no flow change do not create non-trivial lifted edges in $g_i$. Therefore only adjacencies in $G$ such that $f_{i+1} > f_i$ give rise to non-trivial lifted edges in $g_i$, unless they are connected to two duplicated segments.  The number of duplicated adjacencies incident on a segment $s$ which are not connected to a duplicated segment is bounded by $i_s$, hence there are at most $I_i$ adjacencies of this type across the graph. In addition, there are $S_i$ bottom layer adjacencies which were added at random. Hence the total number of non-trivial lifted edges is bounded by $I_i+S_i$. The construction algorithm guarantees that each perfect component in $G$ gives rise to a simple module in $G$, hence, by the upper bound from \citep{avg}, the cost is bounded by $S_i+I_i-l_i$.
\end{proof}

\subsection{Primary extensions}

We say that a valid flow transition $(F_A,F_B)$ is a \emph{lookup} if $C^l_{F_A,F_B} = C^u_{F_A,F_B}$. In this case it is easy to compute the rearrangement cost of the transition. If a transition is not a lookup, then one way to assess its cost is to sample flow sequences with the same initial and final flows that are separated by smaller, more incremental changes. 

A valid sub-sequence $s_2$ of a given valid flow sequence $s_1$ with the same initial and final elements is called a \emph{reduction} of $s_1$, and conversely $s_1$ is an \emph{extension} of $s_2$. 

It is convenient to look at this in terms of flow differences, which are themselves flows. A valid flow transition $(F_A,F_B)$ defines a flow that is the difference between flows $F_A$ and $F_B$, i.e. the flow $\Delta F = F_B - F_A$. Likewise, a valid extension $(F_1, ..., F_{k+1})$ of $(F_A, F_B)$ defines a sequence of nonzero flow differences $\delta F_1, ..., \delta F_k$, where $\delta F_i = F_{i+1}-F_i$, such that: 
$$\sum_{i=1}^{k} \delta F_i = \Delta F$$
We call the multiset of nonzero flows ${\delta f_1, ..., \delta f_k}$ a decomposition of $\Delta F$. 

If, for each $i$, either the flow transition $(F_i,F_{i+1})$ is a lookup or at least simple enough so that we can compute the transition cost in reasonable time, then by sampling decompositions of the overall flow difference $\Delta F$, we can arrange these into extensions of this flow difference, and evaluate these extensions efficiently, keeping track of the cheapest one we find. This provides a sampling strategy for producing plausible explanations for complicated observed flow transitions. To make this work in practice, we introduce the notion of a primary (or near-primary) flow, and demonstrate how to decompose a flow difference $f$ into a sum of primary or near-primary flows. The primary or near-primary flow differences we get are usually a lookup, and if not, the time required to compute their cost is reasonable. 

We first introduce primary flows, which are a generating set of $\mathcal{F}_{\mathbb{Z}}(G)$, then a superset of these, the near-primary flows, which can be sampled ergodically with a simple algorithm.

We measure the elements of $\mathcal{F}_\mathbb{Z}(G)$ with the $L_1$ norm, defined by $\|f\|_1=\sum_{e \in E}|f(e)|$. A non-zero flow $f$ is \emph{primary} if there do not exist two non-zero integer flows $f_1$ and $f_2$ such that $f=f_1+f_2$ and $\|f\|_1=\|f_1\|_1+\|f_2\|_1$.

\begin{theorem}
\label{primary_decomp}
Primary flows are a generating set of $\mathcal{F}_{\mathbb{Z}}(G)$.
\end{theorem}

\begin{proof}
We will decompose a flow $f$ by induction. For $k \ge 1$, we will define a set $S_k$ of $k$ non-zero flows $(f^k_1, \ldots, f^k_k)$ (allowing for duplicate flows), such that $f = \sum_{f' \in S_k}f'$ and $\|f\|_1 = \sum_{f' \in S_k} \|f'\|_1$. At the start, $S_1=(f)$. If there exists $i \le k$ such that $f^k_i$ is not primary, then there exist two non-zero integer flows $f_b$ and $f_c$ such that $f^k_i=f_b+f_c$ and $\|f^k_i\|_1=\|f_b\|_1+\|f_c\|_1$. We define $S_{k+1}=(f^k_1,\ldots,f^k_{i-1},f_b,f_c,f^k_{i+1},\ldots,f^k_k)$. It is straightforward to verify that $S_{k+1}$ fulfills the two necessary equalities. We proceed until no non-primary flows remain. In that case, then $f$ was successfully decomposed as a sum of primary flows. Since the $L_1$ norm of a non-zero flow is necessarily a non-zero integer, the total number $k$ of flows we create in this way is bounded by $\|f\|_1$. The non-zero flow $f$ was therefore decomposed as a finite sum of primary flows.
\end{proof} 

A \emph{valid primary flow sequence} is a valid flow sequence $s=(f_1,\ldots,f_k)$ such that for each flow transition, $(f_{i},f_{i+1})$, its associated \emph{flow change} $(f_{i+1} - f_i)$ is a primary flow. 

\begin{corollary}
Any valid sequence of flows in $\mathcal{F}_{\mathbb{Z}}(G)$ can be extended into a valid primary flow sequence.
\end{corollary}

\begin{proof}
We decompose the flow changes of a valid sequence into primary flow changes using the approach described in Theorem \ref{primary_decomp}. At each step, a flow change $f$ is replaced by two flows $f_1$ and $f_2$ such that a) $f=f_1+f_2$ and b) $\|f\|_1=\|f_1\|_1+\|f_2\|_1$. From equality a), it follows that at every edge $e$ we have $0 < |f(e)| \le |f_1(e)|+|f_2(e)|$. Given equality b) we  have at every edge $e$ $|f(e)| = |f_1(e)|+|f_2(e)|$, hence $f_1$ and $f_2$ have the same sign as $f$. Therefore if $f$ is the flow change between two non-negative genome flows $F_A$ and $F_B$, $F_A + f_1=F_B - f_2$ is also non-negative. The extended flow sequence $(\ldots,F_A, F_A + f_1, F_B,\ldots)$, with flow changes $(\ldots,f_1,f_2,\ldots)$, is also a valid flow sequence.
\end{proof}

By the above result, any flow change $f$ can always be decomposed, not necessarily uniquely, into primary flow changes $(f_i)_{i=1..k}$ such that $f = \sum_{i=1}^k f_i$ where $\|f\|_1 = \sum_{i=1}^k \|f_i\|_1$. This type of decomposition is fast to compute, and useful to create instances of primary flow sequences.

However, these decompositions are intrinsically limited in scope. In particular, the second condition implies that when the flow change $f$ is, say, positive on a given edge $e$, then $\forall i < k, f_i(e) \ge 0$. This forbids edge re-use, i.e. cases where adjacencies are temporarily created, then deleted, or segments temporarily duplicated, then deleted. Edge reuse is common in actual rearrangement histories, including minimum cost histories, so we do not always want to impose this second condition. 

Given a flow $f$, a \emph{primary extension} of $f$ is a set $\{c_1 f_1,  \ldots, c_n f_n\}$, such that $\{c_1, \ldots, c_n\} \in {\mathbb{N}^*}^n$ and $f_1, \ldots, f_n$  are primary flows, such that $f = \sum_{i=1}^{n} c_i f_i$ and the component flows $f_1, \ldots, f_n$ are linearly independent, i.e. no component can be written as a linear combination of the others. Note that since the dimension of the subspace of all flows is at most $\|E\|$, where $E$ is the set of edges in the graph, no primary extension can have more than $\|E\|$ components. Further, since the components $f_i$  are required to be linearly independent, once these are specified for a given $f$, the coefficients $c_i$ are uniquely determined. 

This definition eliminates trivially non-parsimonious cases where a simpler decomposition could be obtained with a strict subset of the component flows. This amounts to forbidding a some cases of homeoplasy, i.e. we allow a simple rearrangement event to happen multiple times, in which case the component is assigned a weight greater than 1, for example in the case of tandem duplications, but we don't allow distinct sets of components to have identical or inverse flow changes, creating a new configuration and then undoing it again for no net effect.

\subsection{Characterising the space of primary flows}

We now demonstrate the ties between primary flows and the space of even length cycles. 

For any weighting $f$ in $\mathbb{R}^E$, its \emph{conjugate weighting} $\hat f$ is the real-valued function on the edges of $E$ defined by:
\begin{eqnarray*}
\hat f: & E & \mapsto \mathbb{R} \\
& e & \rightarrow \left\{ \begin{array}{rl}
f(e) & \mbox{if $e$ is a segment edge} \\
-f(e) & \mbox{otherwise} \end{array} \right.
\end{eqnarray*}

\begin{figure}[htbp]
\begin{center}
\includegraphics[width=13cm]{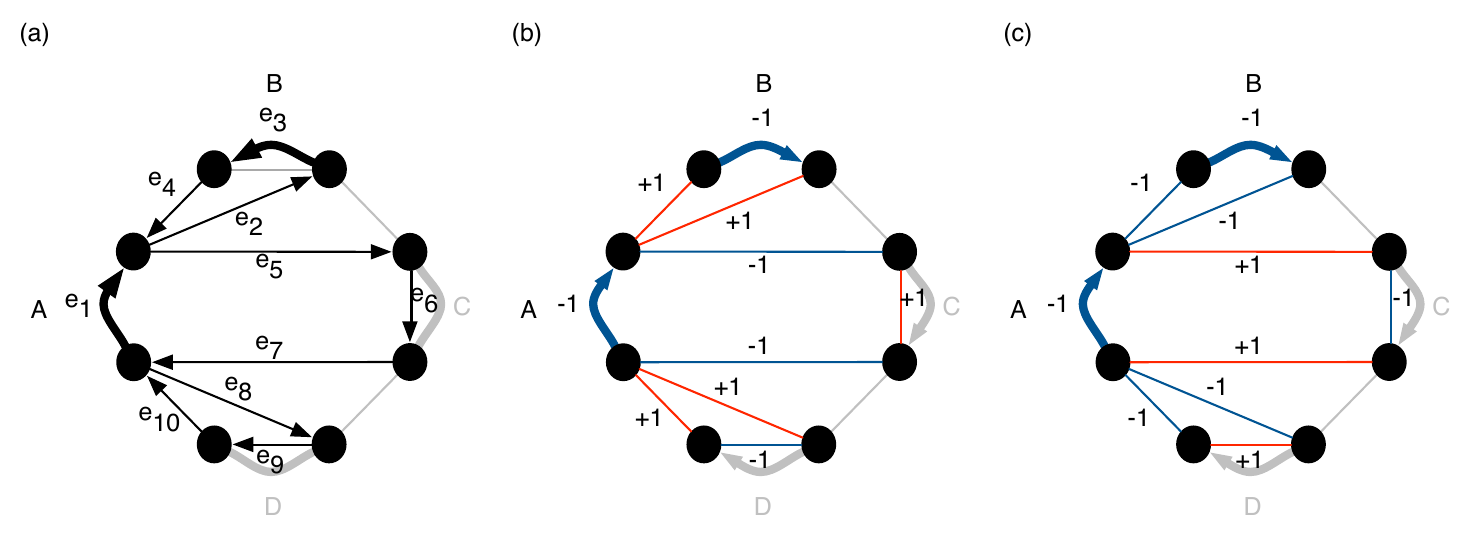}
\caption{(a) An even length cycle traversal, (b) its alternating weighting, and (c) the conjugate of the alternating weighting (i.e. the alternating flow of the cycle traversal). Note how the alternating weighting in (b) respects the conjugate balance condition and the alternating flow in (c) respects the balance condition.}
\label{alternating}
\end{center}
\end{figure}

See Figure \ref{alternating}.c for an example weighting and its conjugate. The conjugate transformation is a linear involution, \emph{i.e.}: $\hat {\hat f}=f$ and $\forall (f_1,f_2) \in \mathcal{F}_{\mathbb{Z}}(G)^2, \forall (p,q) \in \mathbb{R}^2,\widehat{pf_1+qf_2}=p\hat f_1+q\hat f_2$.  

If a weighting is such that the sum of incident weights at every vertex is 0, it is said to satisfy the \emph{conjugate balance condition}. A weighting $f$ is a flow iff $\hat f$ satisfies the conjugate balance condition.

For each edge $e$, let $\delta_e$ be the weighting that assigns 1 to $e$, and 0 to all other edges. The set $\{\delta_e\}_{e \in E}$ forms a trivial spanning set of $\mathbb{Z}^E$.

Given the cycle traversal $t=(v, \{e_i\}_{i =1..2n}) \in V \times E^{2n}$ of an even length cycle $c$ (possibly with edge re-use, possibly with 2 or more consecutive adjacency or segment edges), its associated \emph{alternating weighting} is $w(t)=\sum_{i=1}^{2n}(-1)^{i} \delta_{e_i}$. Because the weights of the edges in the cycle alternate between 1 and -1 along an even length cycle, an alternating weighting satisfies the conjugate balance condition, see Figure \ref{alternating}.b).

The \emph{alternating flow} $f$ of an even-length cycle traversal $t$ is the conjugate of its alternating weighting, \emph{i.e.} $f=\hat{w}(t)=\sum_{i=1}^{2n}(-1)^{i}\hat  \delta_{e_i}$. Conversely, $t$ is a \emph{traversal} of $f$. Because an alternating flow is conjugate to a weighting that satisfies the conjugate balance condition, it satisfies the balance condition defined above, and hence is a flow. See Figure \ref{alternating}.c.

For the next few definitions, we assume that $t=(v, \{e_i\}_{i =1..2n}) \in V \times E^{2n})$ is an even length cycle traversal on $G$.

\begin{figure}[htbp]
\begin{center}
\includegraphics[width=10cm]{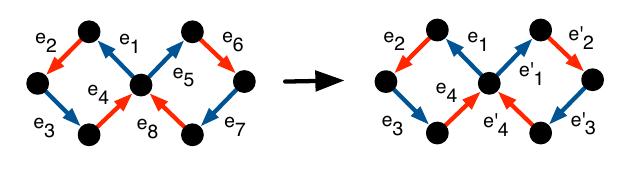}
\caption{On the left, a non-synchronized cycle traversal: edges $e_4$ and $e_8$ coincide, although (8-4) is even. A simple operation transforms it into two synchronized cycles, such that the sum of alternating flows is unchanged.}
\label{synchronized}
\end{center}
\end{figure}

Two distinct indices $i$ and $j$ are said to \emph{coincide} if $e_i$ and $e_j$ end on the same vertex of $G$. The traversal $t$ is \emph{synchronized} if whenever two distinct indices $i$ and $j$ coincide, then $(i-j)$ is odd. See Figure \ref{synchronized} for a counter-example and example. Note that a synchronized traversal cannot have more than two edges that end on the same vertex because for any three integers at least two are even or two are odd, and hence at least one of the pairwise differences among any three integers is even. Hence, a synchronized traversal visits each vertex at most twice. 

\begin{figure}[htbp]
\begin{center}
\includegraphics[width=10cm]{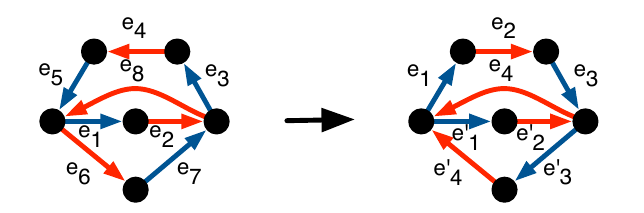}
\caption{On the left, a non-nested cycle traversal: coinciding index pairs $(2,7)$ and $(5,8)$ are such that  $2 < 5 < 7 < 8$. A simple operation transforms it into two nested cycles such that the sum of alternating flows is unchanged.}
\label{nested}
\end{center}
\end{figure}

A set of pairs of integers $P$ is \emph{nested}  if there does not exist $(i_1,j_1), (i_2, j_2) \in P^2$ such that $i_1 < i_2 < j_1< j_2$. The traversal $t$ is \emph{nested} if the set of pairs of coinciding indices of $t$ is nested. See Figure \ref{nested} for a counter-example and an example.

\begin{figure}[htbp]
\begin{center}
\includegraphics[width=10cm]{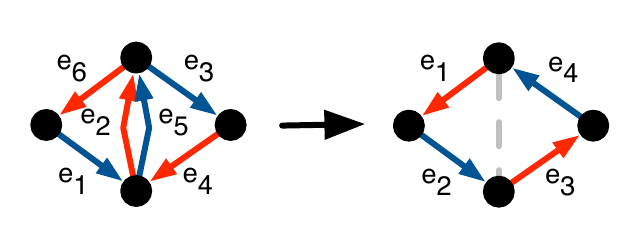}
\caption{On the left, a non-tight cycle traversal: 2 and 5 overlap, although (2-5) is odd. A simple operation transforms it into a tight cycle traversal with identical alternating flow.}
\label{tightFlows}
\end{center}
\end{figure}

The traversal $t$ is \emph{tight} if $e_i=e_j$ implies that $(i-j)$ is even. See Figure \ref{tightFlows} for a counter-example and an example.

\begin{lemma}
\label{l3}
A cycle traversal $t$ is tight iff $\| \hat{w}(t) \|_1 = |t|$, the length of $t$.
\end{lemma}

\begin{proof}
For an edge $e$ of $G$, let $N_e$ be the (possibly empty) set of indices $i$ such that $e_i=e$. Let us suppose that $t$ is tight. This means that $\forall e \in E, \forall (i,j) \in N_e^2, (-1)^i=(-1)^j $.  We then have:

$$\| \hat{w}(t) \|_1 = \sum_{e \in E}  \left| \sum_{i \in N_e} (-1)^i \right| = \sum_{e \in E}  \left| N_e \right| = |t|$$

Conversely, let us suppose that there exists $p$ and $q$ such that $e_p=e_q$ yet $p-q$ is odd. This means that $(-1)^p=-(-1)^q$, we therefore have:

$$
\| \hat{w}(t) \|_1 = \sum_{e \in E}  \left| \sum_{i \in N_e} (-1)^i \right| \le \left[ \sum_{e \in E \setminus \{e_p\}} \left| N_e \right| \right]  + (|N_{e_p}| - 2) \le l-2 < |t|
$$

\end{proof}

\begin{lemma}
\label{nestedsynchtight}
A cycle traversal $t$ that is both nested and synchronized is necessarily tight.
\end{lemma}

\begin{proof}
Assuming by contradiction that $t$ is not tight, there exists two indices $i<j$ such that $e_i = e_j$ and $j-i$ is odd. They cannot be traversed in opposite directions, else $e_i$ and $e_{j-1}$ would coincide at their endpoints, even though $(j-1)-i$ is even, which contradicts synchronization. They cannot either be traversed in the same direction, else $e_i$ and $e_j$ on one hand and $e_{i-1}$ and $e_{j-1}$ on the other would coincide, which contradicts the nesting constraint. 
\end{proof}

A simple cycle is a cycle without vertex reuse. A \emph{cactus graph} is a graph in which any two simple cycles intersect in at most one vertex, which may be called an \emph{intersection point} \citep{harary}. A connected graph is 2-edge connected if for any two distinct vertices A and B, two edge removals are sufficient to disconnect the vertices. It is well-known that a connected graph is a cactus graph iff it is 2-edge connected. A \emph {cutpoint} in a graph is a vertex that if removed splits the connected component in which it resides into two or more connected components called \emph{subcomponents}. 

\begin{lemma}
\label{l4}
Given a simple cycle $C$, one of its traversals $t=(v, \{e_i\}_{i =1..l})$ and a set of properly nested pairs of indices in $[1,l]^2$, merging the vertices at the end of the edges indicated by the index pairs creates a cactus graph.
\end{lemma}

\begin{proof}
We will demonstrate that the graph is 2-edge connected. Let A and B be vertices in the graph. Both $A$ and $B$ can be traced to the merging of two sets of vertices in $C$. Let $I(A)$ be the sets of indices of original vertices merged into $A$, likewise for $B$. Both sets cannot overlap, else they would be identical and $A$ and $B$ would not be distinct. Because of the nesting constraint, no vertex in  $[min(I(B)),max(I(B))]$ can be merged to a vertex outside this interval, therefore cutting the edges at indices $min(I(B))$ and $max(I(B))+1$ breaks the graph into two components. Because all the indices of $I(B)$ belong to the interval, whereas all the indices of $I(A)$ are outside of this interval, $A$ is disconnected from $B$. Thus, the graph is 2-edge connected, in other words it is a cactus graph.
\end{proof}

\begin{figure}[htbp]
\begin{center}
\includegraphics[width=11cm]{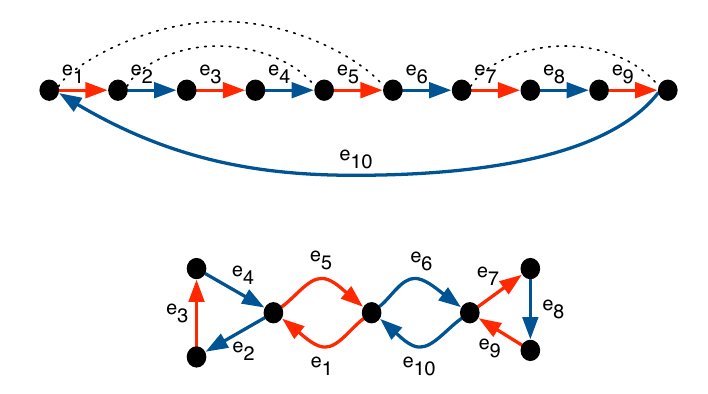}
\caption{Transforming a traversal by merging the end-points of nested coinciding pairs of edges (in this case $\{(1,4),(5,10),(6,9)\}$), shown with dotted lines) into a cactus graph.}
\label{cactus}
\end{center}
\end{figure}

See Figure \ref{cactus} for an example.

\begin{theorem}
\label{t4}
The set of primary flows is identical to the set of alternating flows of nested and synchronized cycle traversals.
\end{theorem}

\begin{proof}
Let $f$ be a primary flow, and $t=(v, \{e_i\}_{i=1..l})$ be a cycle traversal in $G$ that alternates between edges where $f$ is strictly positive and edges where $f$ is strictly negative. We will demonstrate by contradiction that $t$ is nested and synchronized. 

We will first assume that $t$ is not synchronized, i.e. there exist distinct indices $i$ and $j$ such that $i$ and $j$ coincide, yet $(i-j)$ is even. Without loss of generality, we assume that $i<j$. Let $t_1=\{e_{i+1},\ldots,e_j\}$ and $t_2=\{e_1,\ldots,e_i,e_{j+1},\ldots,e_{l}\}$. As $i-j$ is even, both are even length, primary cycles. It follows that $\hat w(t_1) + \hat w(t_2) = \hat{w}(t) = f$, and obviously $|t_1|+|t_2|=l$. Since $t$ alternates between edges where $f$ is positive and negative, $e_i = e_j$ implies $i-j$ is even, hence $t$ is tight. It is obvious that this property is preserved in $t_1$ and $t_2$. Hence, by Lemma \ref{l3}, $ \| \hat w(t_1) \|_1 + \| \hat w(t_2) \|_1 = |t_1| + |t_2| =l $. Therefore $\| f \|_1 =  \| \hat w(t_1) \|_1 + \| \hat w(t_2) \|_1$, hence $f$ could not be minimal. Figure \ref{synchronized} gives an example.

We then assume that $t$ is synchronized, but not nested, i.e. that there exist indices $i<p<j<q$ such that $(i,j)$ and $(p,q)$ are pairs of coinciding indices. Then the traversals $t_1=(v, \{e_1,\ldots, e_i,e_j,e_{j-1}\ldots, e_{p+1},e_{q+1},e_{q+2}\ldots,e_l\})$ and $t_2=(v', \{e_{i+1},e_{i+2}\ldots,e_{p},e_{q},e_{q-1}\ldots,e_{j+1}\})$, where $v'$ is the $(i+1)^{th}$ vertex traversed by $t$, can similarly be used as a decomposition of $f$, verifying that $f$ could not be minimal (Figure \ref{nested}). Thus, $t$ is nested and synchronized.

Let $t=(v, \{e_i\}_{i=1..l})$ be a nested and synchronized cycle traversal, we will demonstrate that its alternating flow $f$ is primary. By Lemma \ref{l4}, because $t$ is nested, the subgraph of G it traverses is a cactus graph. Because it is impossible to find three indices $i,j$ and $k$ such that $(i-j)$, $(j-k)$ and $(k-i)$ are all odd, $t$ can visit any vertex at most twice. Because all the edges in this cactus graph can be traversed in a single cycle traversal, it can be edge-partitioned into a set of simples cycles or chains such that at each vertex at most two components intersect. Because $t$ is synchronized, if $f = f_1 + f_2$ is a decomposition of $f$ and $t_1$ is a traversal of $f_1$, $t_1$ must necessarily switch between edges in different simple cycles whenever it reaches a cutpoint vertex, else the resulting cycle would not be even-length. Thus $t_1$ traverses all of the edges at least once, and is equal to $t$. Therefore $f$ is primary (see Figure 7 for an illustration).
\end{proof}

\subsection{Near-primary extensions}

\begin{figure}[htbp]
\begin{center}
\includegraphics[width=9cm]{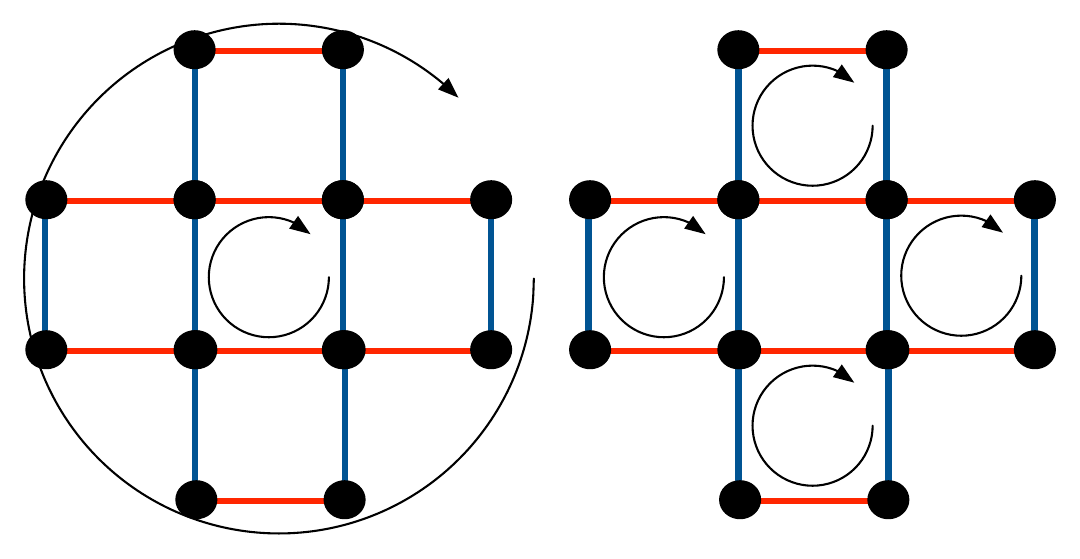}
\caption{In this graph, represented twice, a conjugate flow $\hat f$ is represented with the edge colors. On red edges, $\hat f = 1$, on blue edges $\hat f = -1$. There are exactly two ways of decomposing this conjugate flow as a sum of synchronized and nested alternating weightings, as indicated by the circular arrows. Transforming the righthand decomposition into the lefthand one requires simultaneous modifications of all flows.}
\label{notergodic}
\end{center}
\end{figure}

Although very useful to describe the extensions of a flow $f$, the ergodic exploration of primary extensions is difficult. Figure \ref{notergodic} provides an example where there is no straightforward manner to transform one primary extension into another.

Instead, we focus on a superset of primary flows. A flow derived from a tight and nested traversal is called a \emph{near-primary} flow. Following Lemma \ref{nestedsynchtight}, primary flows are a subset of near-primary flows. To illustrate the difference between the two sets, on Figure \ref{synchronized}, the left traversal is near-primary, the right one is primary. 

A \emph{near-primary extension} of a flow $f$  is a set $(c_1 f_1,  \ldots, c_n f_n)$  for some integer constants $(c_1, \ldots, c_n)$ and near-primary flows $(f_1, \ldots, f_n)$ such that $f = \sum_{i=1}^{n} c_i f_i$ and the component flows $f_1, \ldots, f_n$ are linearly independent. 

\begin{lemma}
The number of possible near-primary extensions of a flow sequence is finite.
\end{lemma}

\begin{proof}
Because a synchronized traversal visits each edge at most twice, there are only a finite number of synchronized traversals, and hence a finite number of near-primary flows. Since the flows in a decomposition $(c_1 f_1, \ldots, c_n f_n)$ of any flow $f$ are linearly independent, $n$ is at most $|E|$, and since we must have $f = c_1 f_1 +  \ldots+  c_n f_n$, the coefficients $c_1, \ldots, c_n$ are uniquely determined by the choice of $f_1, \ldots, f_n$. Hence the number of decompositions is finite.
\end{proof}

This implies that any scoring scheme can be used to construct a valid probability mass function on the set of near-primary extensions of a flow sequence, which constitute a natural set of flow histories.

\subsection{Collapsing equivalent solutions}

Let $K$ be the set of edges with zero overall flow change, and $\Omega$ the set of vertices which are only incident to $K$. To prevent a combinatorial explosion because of equivalent cycles going through $\Omega$, $\Omega$ is collapsed into a universal connector vertex, $\omega$, connected to all other vertices in the graph. In other words, if an alternating flow traversal has edges incident with vertices of $\Omega$, it is represented as going to $\omega$, self-looping on that vertex, then continuing out. This ensures that otherwise equivalent histories are not distinguished because of irrelevant labeling differences between interchangeable vertices. Any alternating flow traversal which self-loops from $\omega$ to itself more than once can be automatically simplified into a simple flow traversal with lesser complexity, avoiding extrapolating useless operations on the edges of $K$. 

Given a valid flow decomposition, if an edge of $K$ is duplicated then deleted, leaving no final CNV change, then the flows of the two events are summed up into a single flow. This constrains partially the space of possible histories, as it precludes a third event from being timed between the two combined events, but it reduces the search space. To mitigate this, this constraint is only applied to the end results of the sampling, as during the sampling stage we allow for events to have non-zero flow over edges of $K$.

After these two transformations, a valid flow decomposition is said to be \emph{collapsed}.

\subsection{Ergodic sampling of near-primary extensions}

If there exist tight and synchronized cycle traversals $t_1=(v, \{e_{1},e_{2}\ldots,e_{2n}\})$ and $t_2=(v, \{\epsilon_{1},\epsilon_{2}\ldots,\epsilon_{2\nu}\})$ such that they start at the same vertex $v$ and overlap over their last $p \ge 0$ edges, but $e_{1} \ne\epsilon_{1}$, we produce a third cycle traversal:

\begin{equation*}
t_3=(v, \{e_{1},e_{2},\ldots,e_{2n-p},\epsilon_{2\nu-p},\epsilon_{2\nu-p-1},\ldots,\epsilon_{1}\}).
\end{equation*}

\begin{figure}[htbp]
\begin{center}
\includegraphics[width=12cm]{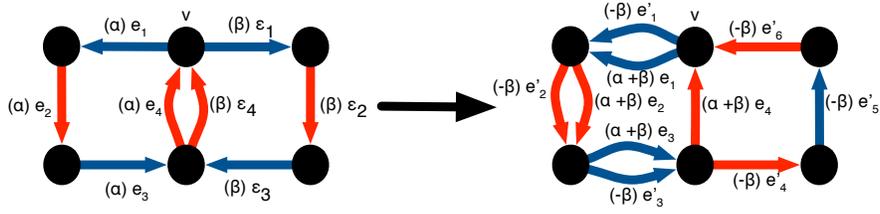}
\caption{Two overlapping flows $t_1$ and $t_2$ transformed by a merge. Here $T = \{t_3\} = \{(v, \{e'_1, ..., e'_6\})\}$. In parentheses are indicated the weightings required to maintain equality of conjugate flow, namely, $\alpha \hat w(t_1) + \beta \hat w(t_2) = (\alpha + \beta) \hat w(t_1) - \beta \hat w(t_3)$}.
\label{merge}
\end{center}
\end{figure}

Subsequently, the traversal $t_3$ is broken up into a set of near-primary cycle traversals $T = \{t'_1, ..., t'_k\}$, using the transformations illustrated in Figures \ref{tightFlows} and \ref{nested}. A \emph{merge} consists in replacing $\{t_1,t_2\}$ with $\{t_1, t'_1, ..., t'_k\}$. Because 
$$\hat w(t'_1) + ... + \hat w(t'_k) = \hat w(t_3) = \hat w(t_1) - \hat w(t_2)$$ we have $\forall (\alpha, \beta) \in \mathbb{R}^2$: $$\alpha \hat w(t_1) + \beta \hat w(t_2) = (\alpha + \beta) \hat w(t_1) - \beta ( \hat w(t'_1) + ... + \hat w(t'_k))$$ Thus the space spanned by $\{t_1, t_2\}$ is spanned by $\{t_1, t'_1, ..., t'_k\}$. See Figure \ref{merge} for an example.

Given two tight cycle traversals $(v, \{e_i\}_{i=1..2n})$ and $(v, \{\epsilon_i\}_{i=1..2\nu})$ that overlap over their first $l$ edges, such that $l$ is odd, we can construct three new cycle traversals:
\begin{eqnarray*}
t'_1 &=& (v', \{E_1,E_2,E_3,e_{l+1}\ldots e_{2n}\}) \\
t'_2 &=&(v', \{E_1,E_2,E_3,\epsilon_{l+1}\ldots \epsilon_{2\nu}\})\\
t'_3 &=& (v, \{e_1,\ldots e_{l},E_3,E_2,E_1\})
\end{eqnarray*}
where $E_1,E_2$ and $E_3$ are adjacency edges such that $E_1$ connects $v'$ to $\omega$, $E_2$ connects $\omega$ to $\omega$ and $E_3$ connects $\omega$ to $v$.

If $l$ is even and greater than zero, then we can construct the following cycle traversals:
\begin{eqnarray*}
t'_1 &=& (v', \{E_1,E_2,e_{l+1}\ldots e_{2n}\}) \\
t'_2 &=& (v', \{E_1,E_2,\epsilon_{l+1}\ldots \epsilon_{2\nu}\}) \\
t'_3 &=& (v, \{e_1,\ldots e_{l},E_2,E_1\})
\end{eqnarray*}
where $E_1$ is a (possibly new) adjacency edge that connects the destination vertex of $e_{l}$, $v'$, to $\omega$, and $E_2$ is a (possibly new) adjacency edge that connects $\omega$ to the starting vertex of $e_{1}$, $v$.

\begin{figure}[htbp]
\begin{center}
\includegraphics[width=12cm]{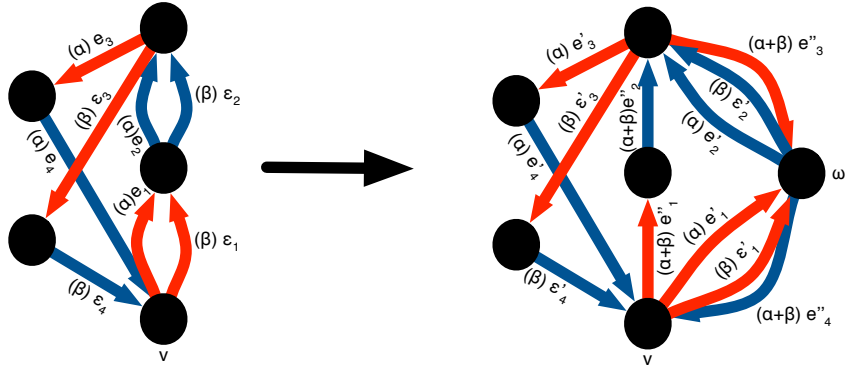}
\caption{Two overlapping flows $t_1$ and $t_2$ transformed by a split into $t'_1=(v,\{e'_1...e'_{2n}\})$, $t'_2=(v,\{\epsilon'_1...\epsilon'_{2\nu}\})$ and $t'_3=(v,\{e''_1...e''_{4}\})$. In parentheses are indicated the weightings required to maintain equality of conjugate flow, namely, $\alpha \hat w(t_1) + \beta \hat w(t_2) = \alpha \hat w(t'_1) + \beta \hat w(t'_2) + (\alpha + \beta) \hat w(t'_3)$}
\label{split}
\end{center}
\end{figure}

This transformation is referred to as a \emph{split}, since it separates out the the edges $e_1, e_2, \ldots e_l$, used in two cycles, into a separate cycle, that uses them only once. See Figure \ref{split} for an example. As in the case of a merge, the resulting traversals are then further broken down into independent near-primary flows. Because 
$$\hat w(t'_1)+\hat w(t'_2)+\hat w(t'_3) = \hat w(t_1) + \hat w(t_2)$$ we have $\forall (\alpha, \beta) \in \mathbb{R}^2$: $$\alpha \hat w(t_1) + \beta \hat w(t_2) = \alpha \hat w(t'_1) + \beta \hat w(t'_2) + (\alpha + \beta) \hat w(t'_3))$$ Thus the space spanned by $\{t_1, t_2\}$ is spanned by $\{t_1, t'_1, ..., t'_k\}$. See Figure \ref{merge} for an example.

\begin{theorem}
The process of applying random merges and splits is ergodic over all the near primary collapsed extensions of a given flow transition. 
\end{theorem}

\begin{proof}
We are given a near-primary collapsed extension $H$ of a flow transition $(f_1,f_2) \in \mathcal{F}(G)^2$, and wish to reach another near-primary collapsed extension $H^\circ$ that also spans $\Delta f = f_2 - f_1$. Because they are both collapsed, $H^\circ$ and $H$ cover exactly the same set of vertices, namely $E \setminus \Omega$.

We choose at random a near-primary component flow $\delta f^\circ$ of $H^\circ$, and one of its tight traversals $t^\circ$. Since $H$ and $H^\circ$ cover the same set of edges, it is possible to greedily define a sequence of flows of $H$ that cover each of the edges of $t^\circ$. The flows can be sequentially merged, since each shares an overlap of at least one vertex with a previously defined flow. Following the greedy merge, they may be greedily split to recover $\delta f^\circ$. It is thus possible to reconstruct any near-primary component flow of $H^\circ$ with splits and merges between the vectors in $H$. 
\end{proof}

\subsection{Fixing invalid flow sequences}

A flow sequence is a sequence of positive flows, from which a sequence of flow changes can be derived. However, a near-primary extension of a positive flow can not necessarily be ordered to define a valid flow sequence because cumulatively adding up the flow changes can produce negative values.

Imposing that each path of the clonal derivation tree define a valid flow sequence would require much computation and possibly break the ergodicity of the method described below. For this reason we preferred to impose a softer constraint, namely increasing the cost of a history showing inconsistencies. For every segment edge which goes from non-negative to negative or from 0 to positive flow, a penalty charge of 2 events is added to the history. This cost corresponds to the cost of temporarily duplicating a region then deleting it. 

\section{Simulations}

\subsection{Measuring accuracies of solutions}

To test our model, we created 90 artificial rearrangement histories with between 2 and 33 mutational events.  Each simulated history begins with a single chromosome composed of 100, 150, or 200 atomic blocks of sequence, and the history is built by subsequently randomly deleting, duplicating and inverting segments of contiguous blocks. The frequency of these events was respectively 70\%, 10\% and 10\% (in the remaining 10\% of cases no change was applied).  Given the final copy number profile and novel adjacencies of the rearranged genome, we sampled possible flow histories using importance sampling based on the above ergodic exploration of the space with ten independent runs of 2,000 iterations each. Each near-primary flow, representing a set of rearrangement events, received a likelihood score equal to the fraction of sampled histories that contain it. So as not to penalise valid flow transitions that are extensions of multiple flow transitions from the simulated history, we counted as true positive any sampled flow transition which could be decomposed as a finite sum of original simulated flow transitions. 

\begin{figure}[htbp]
\begin{center}
\includegraphics[width=7cm]{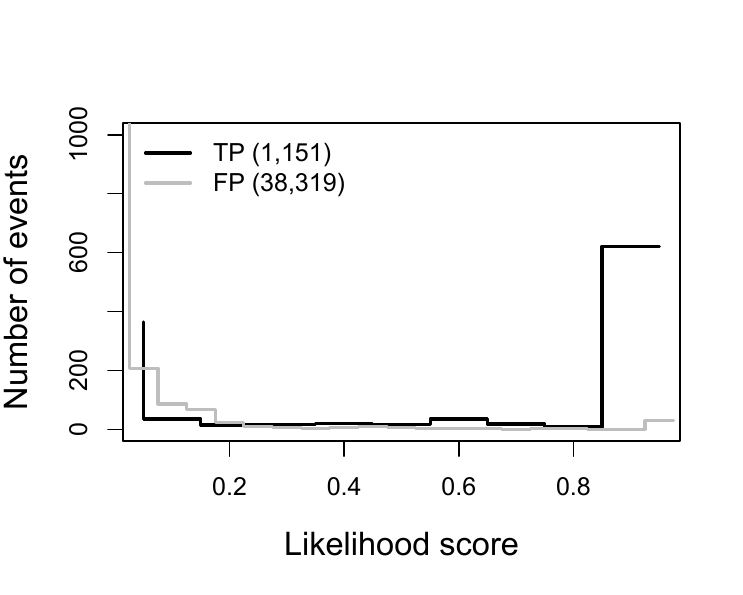}
\caption{Near-Primary flows across 10,000 sampled histories for 90 simulated rearrangement histories were merged and assigned a likelihood score.  Near-Primary flows with high likelihood scores are enriched for simulated events, while near-primary flows with low likelihood scores do not represent the simulated flow history.}
\label{likelihoodHistogram}
\end{center}
\end{figure}

We find that near-primary flow transitions predicted in the sampled histories but not present in the simulated history have low likelihood scores, and that correctly predicted events have high likelihood scores, as shown in Figure \ref{likelihoodHistogram}.  Furthermore, out of the 751 events with a likelihood score greater than 0.5, 699 (93\%) of them are simulated events, so near-primary flows with high likelihood scores are more likely to represent the underlying simulated rearrangement history.

\begin{figure}[htbp]
\begin{center}
\includegraphics[width=12cm]{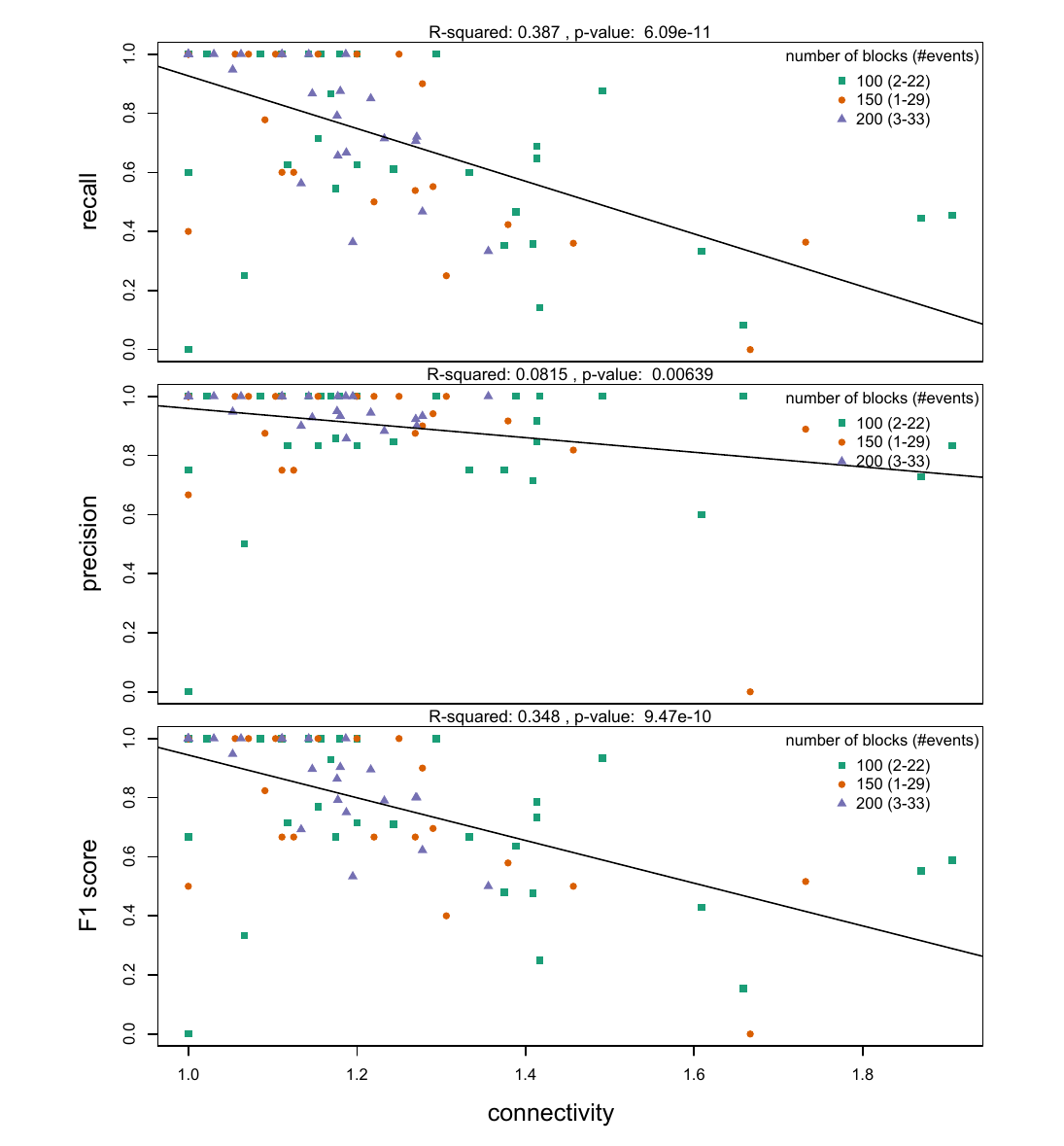}
\caption{We calculated statistics for 90 simulated histories with varying densities of rearrangement events.  The ability to accurately predict a flow history decreases as the breakpoint reuse, the number of edges over the number of nodes, increases.  Most of this loss in accuracy comes from a decrease in sensitivity, or the simulated near-primary flows not being constructed in the sampled histories. }
\label{accuracyScatters}
\end{center}
\end{figure}

After filtering for near-primary flows with likelihood scores greater than 0.5, we determined the recall, precision, and F-score for each of the 90 simulated rearrangement histories.  This is shown in Figure \ref{accuracyScatters} as a function of the connectivity or breakpoint reuse of the simulated flow history (total sum of the norms of the flows divided by the number of nodes in the graph).  The accuracy of a sampled set of histories decreases as the near-primary flows become more connected, either through breakpoint reuse or through nested overlapping events.  Breakpoint reuse introduces coinciding edges in the CN-AVG, as described in figures \ref{merge} and \ref{split}.  These formations have alternative traversals, leading to ambiguity in the flow history.  For trivial histories with a connectivity equal to one, representing no breakpoint reuse, we achieve up to 100\% accuracy, as measured by the F-score.  For histories with connectivity greater than 1, we achieve an average 77\% accuracy, largely due to a drop in sensitivity with increased breakpoint reuse.  This indicates that certain complicated and entangled rearrangement events cannot be reliably segregated from the simulated near-primary flows.  In 25/90 (28\%) of the simulations, there exists a sampled history with a lower cost than the simulated history, so we would not expect all near-primary flows in the simulated history to have the highest likelihood scores in these cases.  

\section{Discussion}

As discussed in the introduction, there have been many attempts to define parsimomious histories given a set of rearranged genomes with deletions and duplications. Because of the inherent complexity of the problem, different models made different simplifying assumptions. The model presented here is particularly flexible, in that it allows arbitrarily complex rearrangement events containing duplications and deletions of arbitrary contiguous regions, not just predefined atomic components or predefined rearrangement transformations. To avoid the ``free lunch'' problem pointed out by \citet{DCJindels}, it does not allow \emph{de novo} insertions. In addition, this is first such model which offers an explicit ergodic sampling procedure. In counterpart, we currently bracket the cost of each flow transition between an upper and lower bound, with the expectation that the two bounds meet for most encountered near-primary flow transitions. Given their constrained size, it should be possible to exhaustively test the cost of near-primary flow transitions when these two bounds diverge. In addition, the collapsing of histories does preclude some event orderings.

Theorem 2 only describes the cost of each epoch independently, and does not allow us to extrapolate the cost of entire histories.  Because the bilayered components of a layered history graph share intermediary layers, defining the structure of one layer constrains the possible layers immediately before and after it. It is therefore sometimes impossible to construct a realisation of a flow sequence such that every bilayered component has a cost below its corresponding upper bound. Determining the cost of entire histories would presumably require reconstructing these histories in their entirety, as we describe in \citep{avg}. However, this explicit approach requires the ancestral and derived genomes to be fully determined, which is generally not the case when working with sequencing data. The loss of precision therefore appears to be necessary to handle the incompleteness of genomic data. As genomic sequencing progresses, it might be possible to obtain near-finished assemblies straight from the data, in which case the model described in \citep{avg} would be more relevant.

From a mathematical standpoint, the set of flows of $G$ is very rich. It bears similarity to the cycle space described by \citet{maclane}. The balance condition guarantees that a positive flow can be decomposed as a set of weighted threads, and that in a flow transition, every created adjacency is compensated by a either a corresponding adjacency deletion or a segment duplication, much like double entry book-keeping. The near-primary flows further provide us with a finite subset of flows, such that all possible realisations of the data can be sampled ergodically. However, unlike models for single base substitutions and balanced rearrangements, this model of evolutionary cost is not a distance function, because it is asymmetric, \emph{i.e.} the distance between two genomes depends on which genome is ancestral.  For example, duplicating a segment then translocating one of its copies generally requires two DCJ operations, yet only one operation (a segmental deletion) is needed to return to the original genome. 

\section{Conclusion}

We presented here a model to efficiently sample the space of possible rearrangement histories in the presence of duplications and deletions. Confronted with the NP-hard problem of inferring parsimonious histories described in \citep{avg} from everyday genomic data, we opted for a simplification of the problem which allows us to adopt an efficient ergodic sampling strategy. Through simulation, we verified that this approach produces exploitable results. In practice, it is capable of accurately discriminating true rearrangement events and sampling their possible orderings. There remain a few open questions, in particular whether it is possible to efficiently compute the rearrangement cost of any primary flow sequence.

\section{Availability}

The code used to do the above simulations and tests is freely available at \textbf{https://github.com/dzerbino/cn-avg}.

\section{Acknowledgements}

We would like to thank the Howard Hughes Medical Institute, Dr. and Mrs. Gordon Ringold, NIH grant 2U41 HG002371-13 and NHGRI/NIH grant 5U01HG004695, and the European Molecular Biology Laboratory (DZ and TB) for providing funding.

\bibliography{BalancedFlows}

\end{document}